\documentclass[usegraphicx,usenatbib,useAMS]{mn2e}
\usepackage{amsmath}
\usepackage{amssymb}
\usepackage{color}
\usepackage{url}
\usepackage{threeparttable}

\newcommand{\lya}{\ifmmode\mathrm{Ly}\alpha\else{}Ly$\alpha$\fi}
\newcommand{\lyb}{\ifmmode\mathrm{Ly}\beta\else{}Ly$\beta$\fi}
\newcommand{\igm}{\ifmmode\mathrm{IGM}\else{}IGM\fi}
\newcommand{\lae}{\ifmmode\mathrm{LAE}\else{}LAE\fi}
\newcommand{\h}{\ifmmode\mathrm{H}\else{}H\fi}
\newcommand{\hi}{\ifmmode\mathrm{H{\scriptscriptstyle I}}\else{}H\,{\scriptsize I}\fi}
\newcommand{\hii}{\ifmmode\mathrm{H{\scriptscriptstyle II}}\else{}H\,{\scriptsize II}\fi}
\newcommand{\cmb}{\ifmmode\mathrm{CMB}\else{}CMB\fi}
\newcommand{\qso}{\ifmmode\mathrm{QSO}\else{}QSO\fi}
\newcommand{\eor}{\ifmmode\mathrm{EoR}\else{}EoR\fi}
\newcommand{\cmmc}{\textsc{\small 21CMMC}}
\newcommand{\cmfst}{\textsc{\small 21CMFAST}}

\title[IGM temperature at $z=8.4$]{Constraints on the temperature of the intergalactic medium at $z=8.4$ with 21-cm observations}

\author[B. Greig et al.] {Bradley~Greig$^{1}$\thanks{E-mail:~bradley.greig@sns.it}, Andrei~Mesinger$^{1}$ and Jonathan C. Pober$^{2}$ \\
$^1$Scuola Normale Superiore, Piazza dei Cavalieri 7, I-56126 Pisa, Italy \\
$^2$Physics Department, Brown University, Providence, RI 02912, USA
}

\begin{document}
\maketitle \begin{abstract}
\noindent
 We compute robust lower limits on the spin temperature, $T_{\rm S}$, of the  $z=8.4$ intergalactic medium (\igm{}), implied by the upper limits on the 21-cm power spectrum recently measured by PAPER--64. Unlike previous studies which used a single epoch of reionization (\eor{}) model, our approach samples a large parameter space of \eor{} models: the dominant uncertainty when estimating constraints on $T_{\rm S}$. Allowing $T_{\rm S}$ to be a free parameter and marginalizing over \eor{} parameters in our Markov Chain Monte Carlo code \cmmc{}, we infer $T_{\rm S}\ge$~3~K (corresponding approximately to 1$\sigma$) for a mean IGM neutral fraction of $\bar{x}_{\hi{}}\gtrsim0.1$. We further improve on these limits by folding-in additional \eor{} constraints based on: (i) the dark fraction in QSO spectra, which implies a strict upper limit of $\bar{x}_{\hi{}}[z=5.9]\leq 0.06+0.05 \,(1\sigma)$; and (ii) the electron scattering optical depth, $\tau_{\rm e}=0.066\pm0.016\,(1\sigma)$ measured by the {\it Planck} satellite.  By restricting the allowed \eor{} models, these additional observations tighten the approximate 1$\sigma$ lower limits on the spin temperature to $T_{\rm S} \ge 6$ K. Thus, even such preliminary 21-cm observations begin to rule out extreme scenarios such as `cold reionization', implying at least some prior heating of the \igm{}.  The analysis framework developed here can be applied to upcoming 21-cm observations, thereby providing unique insights into the sources which heated and subsequently reionized the very early Universe.
\end{abstract} 

\section{Introduction}

Radiation emitted from the first stars and galaxies marked the end of the cosmic dark ages. The UV ionizing photons from these first sources escape into the intergalactic medium (\igm{}), reionizing the pervasive neutral hydrogen fog, bringing about the final major baryonic phase change of the Universe. This epoch of reionization (\eor{}) is rich in information on the formation and evolution of structures in the early Universe, including the nature of the first stars and galaxies and their impact on the \igm{} \citep[see e.g.][]{Barkana:2007p2929,Loeb:2013p2936,Zaroubi:2013p2976}. 

The most promising observational tool for probing the \eor{} physics is the redshifted 21-cm spin--flip transition of neutral hydrogen\citep[see e.g.][]{Gnedin:1997p4494,Madau:1997p4479,Shaver:1999p4549,Tozzi:2000p4510,Gnedin:2004p4481,Furlanetto:2006p209,Morales:2010p1274,Pritchard:2012p2958}. The signal is expressed as the offset of the 21-cm brightness temperature, $\delta T_{\rm b}(\nu)$, relative to the \cmb{} temperature, $T_{\gamma}$ \citep[e.g.][]{Furlanetto:2006p209}:
\begin{eqnarray} \label{eq:21cmTb}
\delta T_{\rm b}(\nu) &\approx& 27x_{\hi{}}(1+\delta_{\rm nl})\left(\frac{H}{{\rm d}v_{\rm r}/{\rm d}r+H}\right)
\left(1 - \frac{T_{\gamma}}{T_{\rm S}}\right) \nonumber \\
& & \times \left(\frac{1+z}{10}\frac{0.15}{\Omega_{\rm m}h^{2}}\right)^{1/2}
\left(\frac{\Omega_{\rm b}h^{2}}{0.023}\right)~{\rm mK},
\end{eqnarray}
where $x_{\hi{}}$ is the neutral fraction, $T_{\rm S}$ is the gas spin temperature, $\delta_{\rm nl}(\boldsymbol{x},z)$ is the evolved (Eularian) overdensity, $H(z)$ is the Hubble parameter, ${\rm d}v_{\rm r}/{\rm d}r$ is the gradient of the line-of-sight component of the velocity and all quantities are evaluated at redshift $z = \nu_{0}/\nu - 1$, where $\nu_{0}$ is the 21-cm frequency.  Thus, the 21-cm signal probes both the ionization and thermal state of the \igm{}, and indirectly also the galaxies which regulate them.  Moreover, as it is a line transition, it promises a 3D view of the early Universe.

Tapping into this physical bounty requires sensitive radio interferometers. First generation experiments such as the Low Frequency Array (LOFAR; \citealt{vanHaarlem:2013p200,Yatawatta:2013p2980})\footnote{http://www.lofar.org/}, the Murchison Wide Field Array (MWA; \citealt{Tingay:2013p2997})\footnote{http://www.mwatelescope.org/} and the Precision Array for Probing the Epoch of Reionization (PAPER; \citealt{Parsons:2010p3000})\footnote{http://eor.berkeley.edu/} seek the statistical characterization of the 21-cm signal through a power spectrum (PS) measurement.  Though no detection has been forthcoming thus far, steady progress is being made. Presently, the tightest upper limits come from the 64-dipole PAPER array, which constrained the spherically averaged PS over $0.1\lesssim$ $k$ $\lesssim 0.4$~$h$/Mpc to be less than $\lesssim500$ mK$^2$. Tomographic maps will have to wait for second-generation instruments such as the Square Kilometre Array (SKA; \citealt{Koopmans:2015:p1})\footnote{https://www.skatelescope.org} and the Hydrogen Epoch of Reionization Array (HERA; \citealt{Beardsley:2014p1529})\footnote{http://reionization.org}.

Even these preliminary observations begin to constrain the state of the early Universe.  From equation (1), we see that, once the \igm{} spin temperature is much hotter than the \cmb{} ($T_{\rm S} \gg T_{\gamma}$), the maximum achievable contrast in the 21-cm signal is of order 10 mK, sourced by order unity fluctuations in the neutral fraction during patchy reionization. However, prior to \igm{} heating when the signal is still in absorption against the \cmb{}, the achievable dynamic range is much larger.  For example, the adiabatically cooling \igm{} can reach values of $\delta T_{\rm b}\sim -200$mK.\footnote{This signal is achieved if the spin temperature is efficiently coupled to the gas kinetic temperature, through the Wouthuysen--Field mechanism \citep{Wouthuysen:1952p4321,Field:1958p1}. Simple scaling relations \citep[e.g.][]{Furlanetto:2006p3782,McQuinn:2012p3773} show that this condition should be satisfied well before the \igm{} heating epoch.}  Although the spatial fluctuations in the \igm{} temperature source a strong  21-cm PS signal ($\sim100$s of mK$^2$; e.g. \citealt{Pritchard:2007p3787, Pacucci:2014p4323}), an even stronger signal would arise during so-called cold reionization \citep{Mesinger:2013p1835}. If reionization proceeded in a very cold \igm{}, the resulting contrast between the cold neutral regions (with $\delta T_{\rm b}\sim -100$ to $-200$ mK), and the ionized regions (with $\delta T_{\rm b}\sim0$ mK) could drive the large-scale 21-cm power to values in excess of $\sim1000$s of mK$^2$ (e.g. \citealt{Parsons:2014p781}).  Such models can already be ruled out by current data.

\citet{Parsons:2014p781} were the first to investigate constraints on `cold reionization' from upper limits on the $z=7.7$ 21-cm PS using the 32 dipole PAPER array. \citet{Ali:2015p4327}, using tighter upper limits on the 21-cm PS at $z=8.4$ from an extended 64 element PAPER array, subsequently obtained improved limits of $T_{\rm S} > 4$~K. Both works only used a simple analytic expression for the 21-cm PS during the \eor{}. An improved analysis was performed in \citet{Pober:2015p4328}, making use of the semi numerical code \cmfst{}\footnote{http://homepage.sns.it/mesinger/Sim} \citep{Mesinger:2007p122,Mesinger:2011p1123} to simulate both the \eor{} and \igm{} heating by X-rays. These authors varied the X-ray efficiency of early galaxies, in order to coarsely sample the \igm{} spin temperature, obtaining lower limits of $T_{\rm S} \geq$ 5 K for neutral fractions between 10 and 85 per cent.

Importantly, in each of these studies, only a single \eor{} model was used.  The large-scale morphology (distribution of cosmic \hii{} patches) depends on the properties of the sources and sinks of ionizing photons.  These can produce large variations in both the shape and amplitude of the 21-cm PS during the \eor{}, even at a fixed redshift and mean neutral fraction (e.g.  fig.~2 of \citealt{Greig:2015p3675}).

In order to explore the impact of the \eor{} morphology on the constraints from PAPER--64, we utilize the recently developed Monte Carlo Markov Chain (MCMC)-based \eor{} analysis tool \cmmc{}\footnote{http://homepage.sns.it/mesinger/21CMMC.html} \citep{Greig:2015p3675}. Using this theoretical framework, we recover improved, robust lower limits on the \igm{} temperature by marginalising over \eor{} models. We then strengthen these limits, by including reionization priors from observations of the dark pixel statistics of high-$z$ quasars \citep{McGreer:2015p3668} and the electron scattering optical depth, $\tau_{\rm e}$ \citep{Collaboration:2p3673}.

The remainder of this paper is organized as follows. In Section~\ref{sec:method}, we outline our analysis methodology, while in Section~\ref{sec:results}, we discuss our improved constraints on the \igm{} temperature. Finally, in Section~\ref{sec:conclusion}, we finish with our closing remarks. Throughout this work, we adopt the standard set of $\Lambda$CDM cosmological parameters: ($\Omega_{\rm m}$, $\Omega_{\rm \Lambda}$, $\Omega_{\rm b}$, $n$, $\sigma_{8}$, $H_{0}$) = (0.27, 0.73, 0.046, 0.96, 0.82, 70~${\rm km\, s^{-1} \, Mpc^{-1}}$), measured from \textit{WMAP}  \citep{Bennett:2013p3136} which are consistent with the latest results from \textit{Planck} \citep{PlanckCollaboration:2014p3099}.

\section{Methodology} \label{sec:method}

In \citet{Greig:2015p3675}, we developed \cmmc{}, an MCMC-based analysis tool enabling the exploration of the \eor{} astrophysical parameter space. To simulate the reionization epoch, we assume a relatively popular, three parameter model which can accommodate a large set of physically motivated \eor{} morphologies. In this section, we summarize the \eor{} model sampled within \cmmc{} and the modifications required to model `cold reionization', deferring the reader to \citet{Greig:2015p3675} for more technical discussions.

\subsection{Modelling reionization within \cmmc{}} \label{sec:21CMMC_reion}

For a given \eor{} parameter set, we simulate the 21-cm signal using the semi numerical simulation code \cmfst{}.  \cmfst{} employs approximate but efficient methods for modelling the 21-cm signal, and is accurate when compared to computationally expensive radiative transfer simulations on scales relevant to 21-cm interferometry, $\geq1$~Mpc \citep{Mesinger:2011p1123, Zahn:2011p1171}. The speed and efficiency of \cmfst{} makes it well suited for MCMC sampling.  Below we outline the basic components of the \eor{} simulation; readers interested in more details are encouraged to read \citet{Mesinger:2007p122} and \citet{Mesinger:2011p1123}.

\cmfst{} produces 3D realizations of the \igm{} density, velocity, source and ionization fields. A cubic volume of the linear density field is evolved using the Zel'dovich approximation \citep{Zeldovich:1970p2023}. The ionization fields are estimated using the excursion-set formalism outlined in \citet{Furlanetto:2004p123}, but modified to operate on the evolved density field \citep{Mesinger:2011p1123}.  Following this prescription, the time-integrated number of ionizing photons are compared to the number of baryons within spherical regions of decreasing radius, $R$. A cell within the simulation volume is then classified as fully ionized if, 
\begin{eqnarray}
\label{eq:ioncrit}
\zeta f_{\rm coll}(\boldsymbol{x},z,R,\bar{M}_{\rm min}) \geq 1,
\end{eqnarray}
where $\zeta$ is the ionization efficiency which describes the conversion of mass into ionizing photons (see Section~\ref{sec:Zeta}) and $f_{\rm coll}(\boldsymbol{x},z,R,\bar{M}_{\rm min})$ is the fraction of collapsed matter within a spherical radius $R$ residing within haloes larger than $\bar{M}_{\rm min}$ \citep{Press:1974p2031,Bond:1991p111,Lacey:1993p115,Sheth:1999p2053}. Any cells not fully ionized (partial ionizations) are smoothed by the minimum smoothing scale of the cell, $R_{\rm cell}$, and their ionization 
fraction from internal sources is set to $\zeta f_{\rm coll}(\boldsymbol{x},z,R_{\rm cell},\bar{M}_{\rm min})$.
 
In this work, we adopt a popular three parameter model to characterize the \eor{}: (i) the ionizing efficiency of high-$z$ galaxies; (ii) the mean free path of ionizing photons; (iii) the minimum virial temperature hosting star-forming galaxies. These \eor{} parameters are somewhat simplistic as they in effect average over redshift and/or halo mass dependences. However, this simple model suffices to describe a broad range of \eor{} morphologies, while at the same time providing a straightforward physical interpretation. In addition to these \eor{} parameters, here we include an additional free parameter: the mean \igm{} spin temperature. To help gain some intuition, below we summarize these four parameters, highlighting how they affect the 21-cm signal.

\subsubsection{Ionizing efficiency, $\zeta$} \label{sec:Zeta}

The ionizing efficiency of high-$z$ galaxies (equation~\ref{eq:ioncrit}) can be expressed as 
 \begin{eqnarray} \label{eq:Zeta}
 \zeta = 30\left(\frac{f_{\rm esc}}{0.2}\right)\left(\frac{f_{\star}}{0.05}\right)\left(\frac{N_{\gamma}}{4400}\right)
 \left(\frac{1.5}{1+n_{\rm rec}}\right)
 \end{eqnarray}
 where, $f_{\rm esc}$ is the fraction of ionizing photons escaping into the \igm{}, $f_{\star}$ is the fraction of galactic gas in stars,  $N_{\gamma}$ is the number of ionizing photons produced per baryon in stars and $n_{\rm rec}$ is the average number of recombinations per baryon in the \igm{}. 
 
 The parameter $\zeta$ mainly serves to speed up/slow down reionization.  Within this work, we take a flat prior over the range $\zeta\in[5,200]$, which results in a range of reionization histories which are in broad agreement with current \eor{} constraints (Greig \& Mesinger, in prep). 
  
\subsubsection{Minimum virial temperature of star-forming haloes, $T^{\rm min}_{\rm vir}$} \label{sec:Mthresh}

The minimum threshold for a halo hosting a star-forming galaxy, regulating processes important for star formation such as gas accretion, cooling and retainment of supernovae outflows, can be defined in terms of its virial temperature, $T^{\rm min}_{\rm vir}$, which is related to its halo mass via, \citep[e.g.][]{Barkana:2001p1634}
\begin{eqnarray}
M_{\rm min} &=& 10^{8} h^{-1} \left(\frac{\mu}{0.6}\right)^{-3/2}\left(\frac{\Omega_{\rm m}}{\Omega^{z}_{\rm m}}
\frac{\Delta_{\rm c}}{18\upi^{2}}\right)^{-1/2} \nonumber \\
& & \times \left(\frac{T_{\rm vir}}{1.98\times10^{4}~{\rm K}}\right)^{3/2}\left(\frac{1+z}{10}\right)^{-3/2}{\rm M}_{\sun},
\end{eqnarray}
where $\mu$ is the mean molecular weight, $\Omega^{z}_{\rm m} = \Omega_{\rm m}(1+z)^{3}/[\Omega_{\rm m}(1+z)^{3} + \Omega_{\Lambda}]$, and $\Delta_{\rm c} = 18\upi^{2} + 82d - 39d^{2}$ where $d = \Omega^{z}_{\rm m}-1$. 
The value $T^{\rm min}_{\rm vir}\approx10^{4}$~K corresponds to the minimum temperature for efficient atomic cooling; however, efficient star formation likely requires more massive haloes, which are able to better retain and reincorporate supernovae driven outflows \citep[e.g.][]{Springel:2003p2176}.

The parameter $T^{\rm min}_{\rm vir}$ affects (i) when (at what redshift) reionization occurs, and (ii) the bias of the galaxies which are responsible.  A higher value of $T^{\rm min}_{\rm vir}$ means that reionization happened later, with more large-scale ionization structure (at a fixed value of the mean neutral fraction; e.g.~\citealt{McQuinn:2007p1665}). Here we assume a flat prior over the log of the virial temperature, within the range  $T^{\rm min}_{\rm vir}\in[10^{4},5\times10^{5}]$~K.  The lower limit corresponds to the atomic cooling threshold and the upper limit is roughly consistent with the host haloes of observed Lyman break galaxies at $z\sim6$--8 \citep[e.g.][]{Kuhlen:2012p1506,BaroneNugent:2014p4324}.

 \subsubsection{Mean free path of ionizing photons within ionized regions, $R_{\rm mfp}$} 

The physical scales on which escaping ionizing photons are able to penetrate into the surrounding \igm{} depends strongly on both the number and properties of the absorption systems (such as Lyman limit and more diffuse systems). Typically below the resolution limits of \eor{} simulations, these systems behave as photon sinks, resulting in a maximum physical scale on which \hii{} regions are capable of growing around the ionizing galaxies. The impact of photons sinks can be crudely interpreted as a maximum horizon for the ionizing photons, which can correspond to the maximum filtering scale in excursion-set \eor{} models.  We denote this scale as $R_{\rm mfp}$, noting that it depends on the time-integrated value of the mean free path through the ionized \igm{} during the \eor{} \citep{Sobacchi:2014p1157}.

The parameter $R_{\rm mfp}$ most strongly impacts the large-scale ionization structure, with the large-scale 21-cm power dropping with decreasing $R_{\rm mfp}$. Here, we adopt a flat prior over the range $R_{\rm mfp}\in[5, 40]$~cMpc, motivated by sub-grid models of inhomogeneous recombinations \citep{Sobacchi:2014p1157}, as well analytic estimates \citep{Furlanetto:2005p4326} and hydrodynamical simulations of the \igm{} \citep{McQuinn:2011p3293,Emberson:2013p4325}.
 
\subsubsection{Mean \igm{} spin temperature, $T_{\rm S}$} \label{sec:Tspin}

In order to recover limits on the `cold reionization' model, in this work, we adopt a global value for the \igm{} spin temperature of the neutral cosmic gas during the \eor{}. This is an oversimplification, as we would expect spatial fluctuations in $T_{\rm S}$ sourced by the clustering of X-ray sources and the corresponding mean free path of the X-rays. However, in the relevant regime when the signal is the strongest (advanced stages of reionization with weak X-ray heating), a uniform $T_{\rm S}$ is a good approximation for most of the \igm{} as the hottest parts of the \igm{} which normally source large spatial fluctuations in $T_{\rm S}$ have already been ionized.  The remaining neutral gas is distant from galaxies and its heating is governed by the harder X-rays with longer mean free paths.  This picture was quantitatively confirmed by \citet{Pober:2015p4328}, who note that ignoring the spatial fluctuations in the spin temperature of the neutral \igm{} only impacts the large-scale 21-cm  power during `cold reionization' at the level of $\sim10$ per cent (compared with the factor of 10 dependence of the large-scale power on the \eor{} morphology; e.g. fig. 2 in \citealt{Greig:2015p3675}).

Within this work we explore the range $T_{\rm S}\in[1.9, 30]$~K. This range encompasses the low \igm{} temperatures required to source a 21-cm signal strong enough to be ruled out by the upper limits in \citet{Ali:2015p4327}.  For reference, the \igm{} temperature in the absence of heating by astrophysical structures at $z=8.4$ corresponds to 1.9~K \citep[\textsc{RECFAST};][]{Seager:1999p4330,Seager:2000p4329}, whereas the \cmb{} temperature at that redshift is 25.6~K.

\subsection{Computing the likelihood with \cmmc{}} \label{sec:pipeline}

The principal data used for this analysis are the PS measurements from PAPER--64 presented in \citet{Ali:2015p4327}.  Here, we summarize the significant properties of their analysis; referring the reader to their work for detailed discussions of the observations and data reduction. \citet{Ali:2015p4327} analysed the complete observational campaign conducted with a 64-element PAPER array in South Africa over the span of 135 days from 2012 November to 2013 March. Their analysis focused on a 10~MHz frequency band centred on 151.5~MHz, corresponding to a 21-cm redshift of $z = 8.4$. The end result is a quoted upper limit on the 21-cm PS of (22.4 mK)$^2$ over the range of $0.15 < k < 0.5$~$h$/Mpc, nearly a factor of four lower (in mK$^2$) than the previous best upper limit of \citet{Parsons:2014p781}. Approximately half of this increased sensitivity comes from a doubling of the number of antennas (64 elements, as opposed to the 32 used in the \citealt{Parsons:2014p781} analysis).  The other half stems from an updated analysis method, with the principal three advances coming from improved redundant calibration using the methods of \citet{Zheng:2014p3923}, fringe rate filtering \citep{Parsons:2015p5564}, and optimal quadratic estimators \citep{Liu:2011p3925,Trott:2012p2834,Dillon:2013p3924,Liu:2014p3465,Liu:2014p3466}.

The final PS measurements of \citet{Ali:2015p4327} correspond to nine data points spanning the range $0.1 < k < 0.5$~$h$/Mpc, and their corresponding 2$\sigma$ errors were obtained from a boot-strapping analysis. Within this work, we consider only data points with 2$\sigma$ errors that are consistent with a null detection, as systematic errors remain in the data and will lead to significant excesses of power. Indeed within \citet{Ali:2015p4327}, the `detections' at low $k$ are reported as likely being a result of foregrounds, whereas for the other `detections' the causes are less certain. As a result, we are left with a total of four useable data points, which we can set as upper limits on the shape and amplitude of the 21-cm PS.

In the introductory paper of \citet{Greig:2015p3675}, we performed a maximum likelihood sampling of the \eor{} parameter space using a $\chi^{2}$ statistic. To incorporate only the upper limits on the 21-cm PS, we retain the $\chi^{2}$ statistic, however, we modify the likelihood computation. If the amplitude of the model PS falls below the observed value, we set our $\chi^{2}$ to zero, resulting in equal likelihoods for the corresponding \eor{} parameters. If instead the model amplitude is above the observation, we compute the standard $\chi^{2}$, using the bootstrapped errors from \citet{Ali:2015p4327}. Such an approach ensures we appropriately down weight \eor{} models producing amplitudes brighter than the PAPER--64 limits, while accepting all models fainter\footnote{This approach differs to that adopted by \citet{Pober:2015p4328}. These authors use all data points as an upper limit, not just those consistent with zero. For each $k$-mode, they then compute the probability of obtaining their model 21-cm PS relative to the observation, and construct the likelihood by determining the joint probability of obtaining all data points for the model 21-cm PS. In doing so, these authors allow slightly higher amplitude 21-cm PS models than our approach; however, the relative differences between the two approaches should be minimal.}.

Finally, we remove the conservative 25 per cent modelling uncertainty adopted in \citet{Greig:2015p3675} when simulating the PS from \cmfst{}. This source of error is negligible when compared with the high upper limits on the 21-cm PS. Furthermore, we remove our conservative $k$-mode cut at $k = 0.15$~Mpc$^{-1}$, which was included to remove Fourier modes contaminated by foregrounds, since this was already accounted for in the PAPER--64 limits.

\subsection{Including observational priors on the \eor{}} \label{sec:priors}

In an effort to further improve the constraints on the \igm{} temperature, we consider two observational priors. First, we consider the strongest available constraints on the tail of the reionization epoch through measurements of the dark-pixel fraction of high-$z$ quasars  \citep{McGreer:2015p3668}. This approach provides model independent constraints on the \igm{} neutral fraction at $z=5.9$ of $\bar{x}_{\hi{}}\leq0.06 + 0.05\,(1\sigma)$, indicative of a completely (or almost nearly) reionized Universe by $z\approx6$. 

Secondly, we consider the measurement of the electron scattering optical depth to the \cmb{}, $\tau_{\rm e}=0.066\pm0.016\,(1\sigma)$ \citep{Collaboration:2p3673}. This corresponds to an instantaneous reionization redshift of $z_{\rm re} = 8.8\substack{+1.7 \\ -1.4}$, close to the observational redshift of the PAPER--64 21-cm PS constraints ($z=8.4$).

\begin{figure*}
	\begin{center}
		\includegraphics[trim = 0cm 0.6cm 0cm 0.5cm, scale = 0.87]{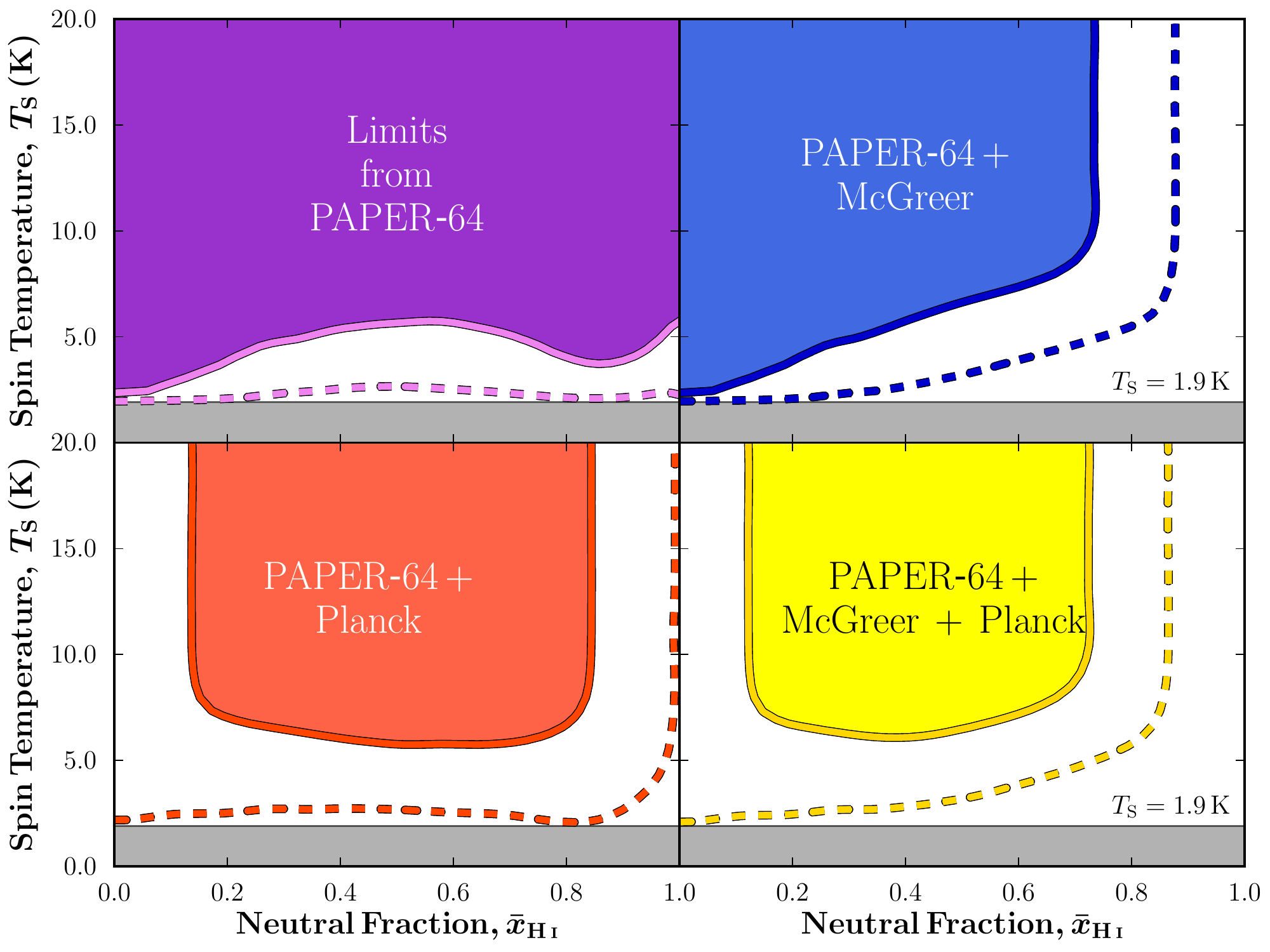}
	\end{center}
\caption[]
        {Joint 2D likelihood surfaces for the \igm{} spin temperature, $T_{\rm S}$ and the \igm{} neutral fraction, $\bar{x}_{\hi{}}$, marginalised over our three morphological \eor{} parameters ($\zeta$, $R_{\rm mfp}$ and $T^{\rm min}_{\rm vir}$) assuming flat (uniform) priors. Shaded regions correspond to $P(\bar{x}_{\hi{}},T_{\rm S}) \geq 0.6$ (approximating 1$\sigma$), while dashed lines correspond to $P \geq 0.15$ (approximating 2$\sigma$). In the top-left panel, we provide the lower limits obtained from only the PAPER--64 constraints on the 21-cm PS. Top right, we include a prior on the \igm{} neutral fraction at $z=5.9$ ($\bar{x}_{\hi{}}\leq0.06 + 0.05$, \citealt{McGreer:2015p3668}) to the PAPER--64 data. Bottom left, the inclusion of a prior on the electron scattering optical depth, $\tau_{\rm e}$ ($\tau_{\rm e}=0.066\pm0.016$, \citealt{Collaboration:2p3673}). Bottom right, our tightest constraints on the \igm{} temperature, including both the $\bar{x}_{\hi{}}[z=5.9]$ and $\tau_{\rm e}$ constraints. In all panels, the grey shaded region represents the lower limit on the \igm{} temperature in the absence of heating by astrophysical structures, $T_{\rm S} = 1.9\,{\rm K}$.}
\label{fig:SpinTSurface}
\end{figure*}

These priors are incorporated into \cmmc{} by computing the 21-cm PS at five different epochs, $z = 6, 7, 8.4, 10$~and~11 for each \eor{} model parameter set in the MCMC, whereas our $\chi^{2}$ estimate is only performed at $z=8.4$. We confirm that this sampling of the reionization history ensures we obtain accurate estimates for $\tau_{\rm e}$.

\section{Results} \label{sec:results}

\subsection{Constraints only from PAPER--64}

In the top-left panel of Fig.~\ref{fig:SpinTSurface}, we report the joint 2D likelihood constraints of the \igm{} temperature and the \igm{} neutral fraction recovered from \cmmc{}. To model the probability surface, we assume a normalized Gaussian distribution whereby each point in the $T_{\rm S}$-$\bar{x}_{\hi{}}$ plane is assigned the probability, $P(\bar{x}_{\hi{}},T_{\rm S}) \propto {\rm exp}(-\frac{1}{2}\chi^{2})$, where the $\chi^{2}$ value is determined by \cmmc{}. In order to obtain this parameter surface, we marginalise over our three morphological \eor{} parameters ($\zeta$, $R_{\rm mfp}$ and $T^{\rm min}_{\rm vir}$) assuming flat (uniform) priors for each across their allowed astrophysical values (see Section~\ref{sec:21CMMC_reion}). In all panels of Fig.~\ref{fig:SpinTSurface}, the shaded regions correspond to likelihoods of $P(\bar{x}_{\hi{}},T_{\rm S}) \geq 0.6$; we consider the regions in white, corresponding to $P(\bar{x}_{\hi{}},T_{\rm S}) \leq 0.6$, to be disfavoured by the data\footnote{Since the observation only consists of an upper limit on the 21-cm PS, the values of $T_{\rm S}\rightarrow \infty$ are allowed with relatively equal likelihood (see equation 1).  Therefore, constructing standard confidence limits on $T_{\rm S}$ (defined to enclose a fixed area of the probability density function) is sensitive to the chosen range of the $T_{\rm S}$ priors.  We instead choose to present our results as contours of equal likelihood, which are well defined, although the exact choice of the threshold value is arbitrary.  The fiducial choice of $P \geq 0.6$ is loosely motivated by the probability of a normalized Gaussian distribution at 1$\sigma$. For illustrative purposes, in all panels in Fig~\ref{fig:SpinTSurface}, we additionally highlight an adopted threshold of $P \geq 0.15$ corresponding to the probability of a normalized Gaussian at 2$\sigma$.}

From the top-right panel, we see that the current PAPER--64 limits constrain the \igm{} temperature to be $T_{\rm S} \gtrsim 5$~K  at greater than 60 per cent confidence, for an \igm{} neutral fraction between 30 and 65 per cent.  These reduce to $T_{\rm S} \gtrsim 3$~K for a broader range in the \igm{} neutral fraction of $\bar{x}_{\hi{}} \gtrsim 10$~per cent.  

Qualitatively, the shape of the allowed $T_{\rm S}$-$\bar{x}_{\hi{}}$ parameter space is almost identical to that presented in \citet{Pober:2015p4328}.  As discussed in that paper, this shape arises from the fact that the spin temperature term in equation (1) can roughly act as a multiplicative factor of the 21-cm PS during the \eor{} under the usual $T_{\rm S} \gg T_{\gamma}$ assumption.  This PS peaks during the midpoint of reionization, and in that regime the \igm{} does not need to be as cold to exceed the \citet{Ali:2015p4327} upper limits.

Quantitatively however, our lower limits on the \igm{} spin temperature are reduced by at least a factor of two compared to those in \citet{Pober:2015p4328}. It is difficult to pin down all causes for this discrepancy, owing to the aforementioned different approaches in constructing the likelihood. Nevertheless, a relaxed lower limit is consistent with our naive expectations. The single \eor{} model used in \citet{Pober:2015p4328} results in sizeable ionization structure on large scales.  On the other hand, our \eor{} parameter space additionally includes models with substantially less large-scale ionization structure (compared at a fixed neutral fraction; see fig. 2 in \citealt{Greig:2015p3675}), mostly sourced by smaller mean free paths of ionizing photons (e.g. \citealt{Sobacchi:2014p1157}). Therefore, by marginalizing over all \eor{} models within our MCMC framework, we would expect a loosening of the lower limits owing to the broadening of the recovered probability distribution for the \igm{} temperature.

\subsection{Constraints including additional \eor{} priors}

In the second panel of Fig.~\ref{fig:SpinTSurface}, we include constraints on the end of reionization at $z\approx6$ from \citet{McGreer:2015p3668}. We find this drastically reduces the allowed parameter space,  disfavouring models with $\bar{x}_{\hi{}} > 0.75$. This behaviour is straightforward to interpret. To produce a $\sim 10$ per cent neutral \igm{} by $z\approx6$ it would be very difficult to have a large $\bar{x}_{\hi{}}$ at $z=8.4$, requiring unphysically-rapid reionization histories.

Next, we consider the inclusion of the $\tau_{\rm e}$ prior from \textit{Planck}, shown in the bottom-left panel of Fig.~\ref{fig:SpinTSurface}. The \textit{Planck} measurement favours a midpoint of reionization close to the same redshift as the PAPER--64 measurement (albeit with broad error bars).  This disfavours models with either too low or too high a neutral fraction, which are precisely those in which the signal is least sensitive to $T_{\rm S}$.   Thus our lower limits on $T_{\rm S}$ improve with respect to the top panels.

Finally, in the bottom right panel, we provide our lower limits combining both observational priors. Owing to the complimentary nature of the two priors, as discussed above, we now recover our most stringent lower limits on the \igm{} temperature of $T_{\rm S} \gtrsim 6$. Even with a lower limits of $T_{\rm S} \approx 6$~K, we can begin to place constraints on reionization scenarios such as `cold reionization'. As this temperature is above that of the \igm{} in the absence of heating by astrophysical structures (1.9~K), this implies that the \igm{} must have undergone some level of heating, ruling out a truly `cold' (unheated) \igm{}. However, the relatively small amount of heating required does not yet begin to strongly constrain the physical processes that could be responsible (e.g. \citealt{Pacucci:2014p4323}).

Nevertheless, this approach highlights the utility of current 21-cm experiments. As more data are acquired, with better characterized systematics, these upper limits on the 21-cm PS amplitude should reduce, tightening the lower limits on the \igm{} temperature.  Alternately, should a high 21-cm PS actually be detected, the implied lack of X-ray heating would have significant consequences for the star formation physics of the first galaxies.

\section{conclusion} \label{sec:conclusion}

One of the major astrophysical goals for the forthcoming decade is the statistical measurement of the 21-cm PS from the \eor{}. In probing the phase change from a neutral to ionized \igm{}, the \eor{} will yield rich information on the formation, growth and evolution of the first stars and galaxies and their influence on the \igm{}. For this reason several, dedicated first-generation radio instruments have been constructed, to be followed by a wave of second-generation experiments with significantly larger collecting areas and sensitivity.

Although these experiments have yet to provide a detection, even current upper limits on the amplitude of the 21-cm PS can begin to constrain the physics of the \eor{}. In the absence of X-ray heating (or any other heating mechanism), the neutral \igm{} should remain `cold', well below the \cmb{} temperature.  If reionization proceeds in such a cold \igm{}, the resulting contrast between cosmic ionized patches (with $\delta T_b \sim 0$ mK) and the cold, neutral patches (with $\delta T_b \sim -100$ mK) would drive the 21-cm signal to values in excess of current upper limits. Recently, both \citet{Ali:2015p4327} and \citet{Pober:2015p4328} used this fact to place limits on the \igm{} temperature during reionization, based on the 21-cm PS upper limits with PAPER--64.

In this work, we improve upon these existing analyses in several ways. Using a modified version of the MCMC-based \eor{} analysis tool \cmmc{} \citep{Greig:2015p3675} we better sample the allowed astrophysical parameter space, improving on the fixed grid approach of \citet{Pober:2015p4328} and the analytic `toy' model of \citet{Ali:2015p4327}. This is achieved by allowing the mean \igm{} spin temperature, $T_{\rm S}$ to be a free parameter rather than varying the X-ray efficiency to recover a mean $T_{\rm S}$ as in \citet{Pober:2015p4328}. Most importantly, by employing \cmmc{}, we marginalize over a broad range of \eor{} models and morphologies. Both previous analyses assumed a single \eor{} model, thereby neglecting the fact that the \eor{} morphology can impact the amplitude of the 21-cm PS by up to an order of magnitude at a fixed neutral fraction (e.g. \citealt{Greig:2015p3675}).

From the PAPER--64 upper limits alone, we can recover lower limits on the \igm{} temperature at $z=8.4$ of $T_{\rm S} \gtrsim 5$~K at greater than 60 per cent confidence for an \igm{} neutral fraction between $30 < \bar{x}_{\hi{}} < 65$~per cent, which reduces to $T_{\rm S} \gtrsim 3$~K when $\bar{x}_{\hi{}} > 10$~per cent. We further tighten these constraints by including \eor{} priors. Folding-in priors on the tail end of the reionization epoch ($\bar{x}_{\hi{}}\leq 0.06+0.05$ at $z=5.9$; \citealt{McGreer:2015p3668}), as well as the integral constraint of $\tau_{\rm e}=0.066\pm0.016$ (\citealt{Collaboration:2p3673}), we obtain lower limits of $T_{\rm S} \gtrsim 6$~K at greater than 60 per cent confidence. While not completely ruling out `cold reionization' models, our results confirm that at least some level of prior \igm{} heating occurred. As both the quality and volume of available 21-cm data continues to increase, our framework will be able to provide unique insights into the high-energy processes inside the first galaxies.

\section*{Acknowledgements}

This project has received funding from the European Research Council (ERC) under the European Union's Horizon 2020 research and innovation programme (grant agreement No 638809 -- AIDA). J.C.P. is supported by an NSF Astronomy and Astrophysics Fellowship under award AST-1302774.

\bibliography{Papers}

\end{document}